\documentclass[a4paper,twoside]{article}
\newcommand{\affil}[1]{$^{\rm #1}$}
\usepackage[authoryear]{natbib}
\bibpunct{(}{)}{;}{a}{}{,}
\usepackage{graphicx}
\usepackage{latexsym}
\usepackage{amsmath}
\usepackage{amssymb}
\usepackage{amstext}
\usepackage{color}
\usepackage{psfrag}
\date{} %Please leave the date blank
\def\k{km s$^{-1}$}
\def\ks{km s$^{-1}$~}
\def\d{$^\circ$}
\def\m{$^\prime$}
\def\sp{$^{\prime\prime}$}
\def\s{$^{\prime\prime}$~}
\def\hh{$^{\mathrm h}$}
\def\mm{$^{\mathrm m}$}
\def\ss{$^{\mathrm s}$}
\def\cm3{cm$^{-3}$}

\def\12{$^{12}$CO}
\def\msol{M$_\odot$}

\title{\large\bf\flushleft An Atomic and Molecular Study of the Interstellar Medium Around
the SNR RCW 103}
\author{\parbox{\textwidth}{\flushleft
\vspace{-0.5cm}
{\it S. A. Paron\affil{A,D}, E. Reynoso\affil{A}, C. Purcell\affil{B}, G. Dubner\affil{A} and A. Green\affil{C}}\\
\vspace{0.4cm}
{\small \affil{A}\,Instituto de Astronom\'{\i}a y  F\'{\i}sica del Espacio (IAFE),
CC 67, Suc. 28, 1428 Buenos Aires, Argentina}\\
{\small \affil{B}\,University of New South Wales, Australia}\\
{\small \affil{C}\,School of Physics, University of Sydney, Australia}\\
{\small \affil{D}\,Email: sparon@iafe.uba.ar}}}

\begin{document}
\twocolumn[
\begin{changemargin}{.8cm}{.5cm}
\begin{minipage}{.9\textwidth}
\vspace{-1cm}
\maketitle
%
%
%%%%%%%%%%%%%     ABSTRACT    %%%%%%%%%%%%%
%Abstract of no more than 200 words here.
\small{\bf Abstract:}
We report on the detection of HCO$^{+}$ and $^{12}$CO emission in the
rotational transition J=1-0 in the vicinity of the shock front at the southern border 
of the supernova remnant RCW 103, where previous infrared observations suggest 
an interaction with a molecular cloud. The observations were carried
out with the Australian Millimeter Radiotelescope at Mopra. We observed a
depletion of HCO$^{+}$ behind the supernova shock front. In addition, we studied
the interstellar medium over an extended region towards RCW 103 based on
archival $\lambda$ 21 cm HI line observations from the Australia Telescope
Compact Array (ATCA) and the Parkes Telescope. No atomic gas is observed in
emission in coincidence with the molecular feature. This absence is interpreted
in terms of self absorption processes.

%%%%%%%%%%%%%     KEYWORDS    %%%%%%%%%%%%%

% Please write all keywords in lower case. PASA uses the
% standard list of subject headings adopted by The Astrophysical Journal
% and available from http://www.journals.uchicago.edu/ApJ/keywords_text.html.
% Keywords are separated by em-dashes, i.e. ---

%%%%%%%%DO NOT EDIT%%%%%%%%%%%%
\medskip
\medskip
\end{minipage}
\end{changemargin}
]
\small
%%%%%%%%EDIT FROM HERE%%%%%%%%%%%%

\section{Introduction}

Supernova remnants (SNRs) mark the catastrophic death of massive stars, and 
they modify irreversibly the surrounding medium into which the shock front is 
expanding. Roughly half of all Galactic SNRs are located near large molecular 
clouds \citep{huang}, presumably the parent clouds of the massive stellar 
progenitors. When an SNR interacts with a molecular cloud, the shock wave 
depletes and excites molecular species, significantly altering the local 
chemistry.

RCW 103 (G332.4-0.4) is a Galactic SNR that shows clear evidence of an 
interaction with a molecular cloud near its southern limb. At radio 
wavelengths, this SNR appears as an almost complete circular shell with a 
diameter of $8^\prime$ \citep{caswell80}. Optical filaments are seen towards 
the brightest regions of the radio shell \citep{vanden,ruiz}. Based on HI 
absorption measurements, \citet{caswell75} estimated a distance of 3.3 kpc for 
RCW 103. Using this distance, \citet{carter} studied the expansion of the 
optical filaments and calculated an age for the remnant of about 2000 years. 
In X-rays, RCW 103 presents a structure compatible with shell SNRs in the young 
double-shock stage of their evolution \citep{cheva}. 

Figure 1 shows a radio continuum image of RCW 103 at 1.4 GHz from \citet
{reyno04} (left) and a soft X-ray image (right) extracted from the {\it 
Chandra} Supernova Remnant Catalog. Good concordance is observed between the 
non-thermal radio emission and the soft X-ray emission. A compact central 
object (CCO) has been detected in hard X-rays \citep{got} near R.A. 16\hh 17\mm 
35\ss , dec. $-$51\d~02\m~25\s~(J2000) with no detected radio counterpart. 
There are strong indications that this CCO is part of a binary system 
\citep[e.g.,][]{becker}. 

Observations in the 2.122 $\mu$m line of the H$_{2}$ and other near infrared 
lines \citep{burton,oliva}, suggest that the remnant is interacting with 
molecular material on the southern side. Burton and Spyromilio conclude that 
the H$_{2}$ spectrum of RCW 103 is typical of dense gas ($\geq$ 10$^{5}$ 
cm$^{-3}$) collisionally excited by low velocity (10 - 20 \ks) shocks. In a 
recent paper, \citet{cheva2005} suggests that RCW 103 is not embedded in a 
molecular cloud, but rather that the blast wave is interacting with previous 
circumstellar material swept up by the progenitor star.

In a search for shocked molecular gas near the southern border of RCW 103 we 
carried out observations in two molecular lines: $^{12}$CO J=1--0 and HCO$^{+}$ 
J=1--0. The first one is the brightest molecular line, very sensitive to the 
presence of molecular gas. Its detection confirms the existence of molecular 
concentrations on the southern side of RCW 103. HCO$^{+}$ emission is enhanced 
at high densities and this molecule is particularly sensitive to shock emission,
providing a good tracer of the shock location. However, this line is generally
quite weak. 

The interstellar HCO$^{+}$ ion has been extensively studied in the past. In 
well-shielded regions, where interstellar radiation plays a minor role 
(molecular regions), the dominant reactions that produce HCO$^{+}$ are the 
following: H$_{2}$ + {\it Cosmic Rays} $\rightarrow$ H$^{+}$ + H + e$^{-}$, 
then H$^{+}$ + H$_{2}$ $\rightarrow$ H$_{3}$$^{+}$, and finally H$_{3}$$^{+}$ 
+ CO $\rightarrow$ HCO$^{+}$ + H$_{2}$. There is much controversy in the 
literature about the way in which shocks modify the HCO$^{+}$ chemistry. 
\citet{iglesias} showed that HCO$^{+}$ must be depleted behind slow shocks. 
However, HCO$^{+}$ observations towards IC443~showed an abundance enhancement 
of this molecule behind the supernova shock front \citep[e.g.,][]{dick}. 
\citet{elitzur} proposed that this enhancement resulted from increased 
ionization due to cosmic rays trapped by shock-compressed magnetic fields. 
However, subsequent observations of IC443 \citep{ziurys,van} showed that the 
HCO$^{+}$ abundance in the shocked gas is not significantly enhanced compared 
with pre-shocked values.

In this paper we investigate the interplay between RCW 103 and its surrounding 
medium based on observations of $^{12}$CO and HCO$^{+}$. A complementary 
analysis of HI $\lambda$ 21 cm data is also presented.

\begin{figure}
\begin{center}
\includegraphics[scale=0.45, angle=0]{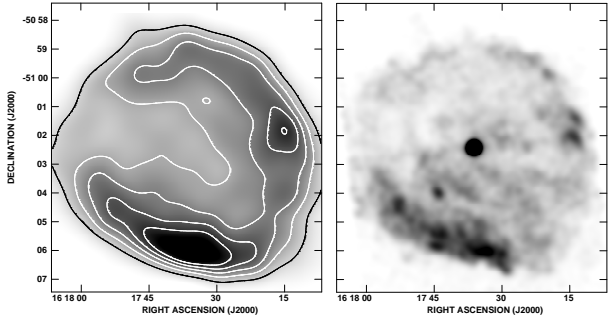}
\caption{{\it Left:} Non-thermal radio emission from RCW 103 at 1.4 GHz 
\citep{reyno04}. The contour levels are 0.1, 0.2, 0.3, 0.4 and 0.5 Jy 
beam$^{-1}$. {\it Right:} Soft X-ray emission in the energy band 1.6-3.0 keV 
({\it Chandra} Supernova Remnant Catalog)}
\end{center}
\end{figure}

\begin{figure}[t]
\begin{center}
\includegraphics[scale=0.4, angle=0]{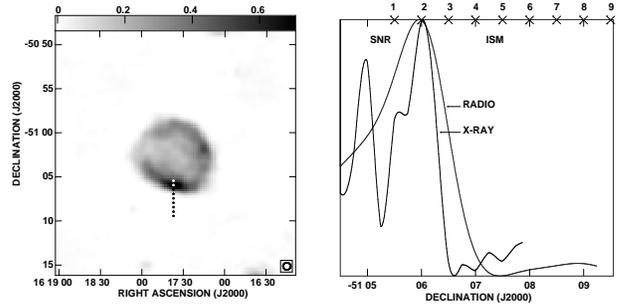}
\caption{{\it Left:} Radio continuum emission at 1.4 GHz in an extended field 
around RCW 103 \citep{reyno04}. The nine dots towards the South of the 
SNR indicate the pointings for the molecular line observations. The grayscale 
is in units of Jy beam$^{-1}$. The beam, 50\s$\times$~50\s, is plotted in the 
bottom right corner. The rms noise level of this image is 5.5 mJy beam$^{-1}$. 
{\it Right:} A line profile at constant R.A. (16\hh17\mm37\ss) for 1.4 GHz 
radio continuum and X rays (1.6-3.0 keV, from {\it Chandra} Supernova Remnant 
Catalog). The crosses with numbers on top indicate the pointings of the 
molecular line observations. }
\end{center}
\end{figure}

\section{Observations}

\subsection{Millimetre Line Observations}

We observed the J\,=\,1\,--\,0 rotational transitions of $^{12}$CO
($\nu$ = 115\,GHz) and HCO$^+$ ($\nu$ = 89\,GHz) using the 22-m Mopra
antenna, near Coonabarabran, NSW, Australia. Spectra were recorded in 
position-switching mode, with the telescope alternating between the 
target and an emission-free reference position. The back-end consisted 
of an auto-correlator, configured to have a bandwidth of 64\,MHz 
divided into 1024 channels, which provided a velocity resolution of 
$\sim$\,0.2\,kms$^{-1}$ over a usable bandwidth of $\sim$\,120\,kms$^{-1}$. 
The pointing accuracy was checked before each mapping run by observing a 
nearby SiO maser, and is estimated to be better than 10\,$''$. Nine pointings, 
at constant right ascension (R.A. = 16\hh17\mm37\ss) were observed towards the 
southern rim of RCW 103 with a resolution of $\sim$34\,$''$. Pointings were 
spaced at 30\s intervals in declination, from $-$51\d05\m30\s to
$-$51\d09\m30\s (Figure 2, left). Also in Figure 2 (right), we 
include a radio continuum and an X-ray emission profile at constant Right 
Ascension (R.A. = 16\hh17\mm 37\ss) to show the SNR shock front position. In 
this picture, the SNR outer shell peaks at dec. $\sim$ $-$51\d~6\m, and the
intensity decreases towards the shell centre (smaller negative declinations)
on the left side of the plot. The crosses with numbers at the top indicate 
the positions of the molecular line observations, and correspond to the 
numbered profiles in Figure 3. Pointings 1 and 2 lie inside the SNR, pointing
3 is at the rim (as seen in radio wavelengths), and the rest probe gas external 
to the remnant.

\subsection{HI Observations}

The $\lambda$ 21 cm HI data used in the present paper are described in detail 
in \citet{reyno04}. These observations were carried out using the Australia 
Telescope Compact Array (ATCA) in the 750A array of a 1\d~x 1\d~region around 
RCW 103, during a 12 h run on 2002 January 22. The baselines in this array 
range from 76.5 to 735 m. A second observation of 12 h with the telescope in the 
EW 367 array (baselines from 46 to 367 m) took place on 2002 April 1. The 
angular resolution of the combined data is 50\sp. A correlator configuration of 
1024 channels was used with a total bandwidth of 4 MHz centered at 1420 MHz. 
The velocity resolution at this frequency is 1 \k. Continuum data were obtained 
simultaneously with a bandwidth of 128 MHz centered at 1384 MHz. Single dish 
data from the Southern Galactic Plane Survey (SGPS) \citep{mcclure05} in the 
$\lambda$ 21 cm HI line were added in order to recover the contribution from 
lower spatial frequencies. 

\section { Results }

\subsection{Molecular Gas}

Figure 3 displays all the observed profiles in the J=1-0 transition of 
$^{12}$CO and HCO$^{+}$ in the velocity range between $-$60 \k ~and $-$20 \k. 
In this velocity interval, the $^{12}$CO spectra have several components, with 
the brightest one centered near $-$48 km~s$^{-1}$. The HCO$^{+}$ spectra are, 
as expected, simpler with only one conspicuous feature centered at a velocity 
varying between $-$45 and $-$50 km~s$^{-1}$ along the pointings, broadly in 
coincidence with the most intense $^{12}$CO peak (all kinematical velocities 
in this paper are referred to the LSR). 

In Table 1 we present the observed parameters of the molecular line components 
peaking near $-$48 km~s$^{-1}$. The peak temperatures (T$_{peak}$), central 
velocities (V$_{c}$) and velocity widths ($\Delta$v) were obtained from 
Gaussian fits to the main components of the spectra. The $^{12}$CO central 
velocities correspond to the most intense peaks. The rms noise is~ $\sigma_{\rm 
CO}\sim$~0.15~K and~ $\sigma_{\rm HCO^{+}}\sim$~0.02~K. The last column 
represents the ratio between the HCO$^{+}$ and $^{12}$CO peak temperatures. 

\begin{figure*}
\centering
\includegraphics[width=13cm]{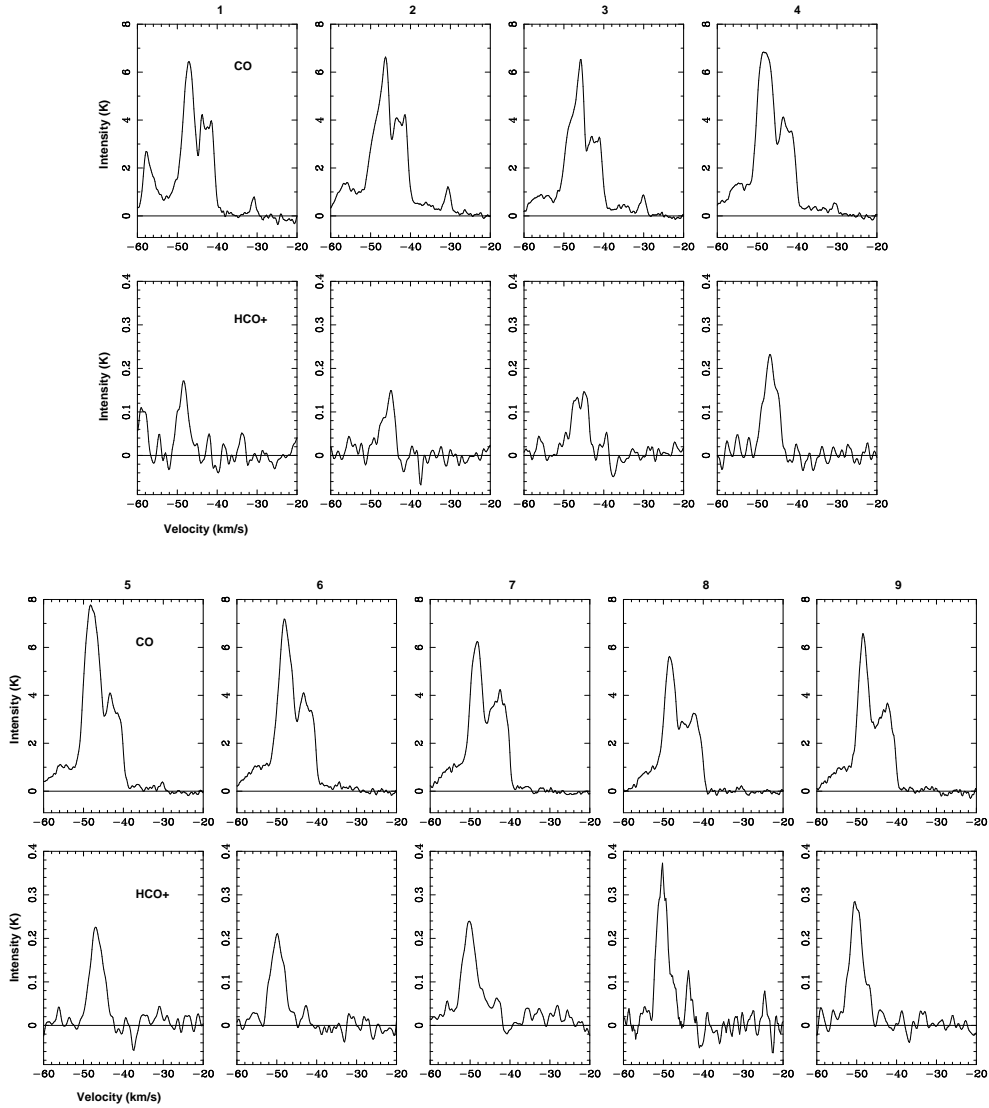}
\bigskip
\caption{The $^{12}$CO and HCO$^{+}$ spectra are displayed in pairs with the 
$^{12}$CO profiles in the upper panel and HCO$^{+}$ in the lower one for every 
pointing, numbered as in Figure 2. The plotted velocity range is [$-$60 \k, 
$-$20 \k] for both molecular transitions. The temperature scales vary from $-$1 
to 8 K for $^{12}$CO and from $-$0.1 to 0.4 K for HCO$^{+}$.}
\end{figure*}

\begin{table*}
\caption{Observed and derived parameters.}
\begin{tabular}{cccccccccccc}
\hline
     \multicolumn{1}{c}{Pointing}&&  
     \multicolumn{2}{c}{T$_{peak}$ (K)}&&
     \multicolumn{2}{c}{V$_{c}$ (km s$^{-1}$)}&&
     \multicolumn{2}{c}{$\Delta$v (km s$^{-1}$)}&&
     \multicolumn{1}{c}{T$^{\rm HCO^{+}}$/T$^{^{12}\rm CO}$ ($\times 10^{-2}$)} \\
 \hline 
    &&$^{12}$CO & HCO$^{+}$ && $^{12}$CO &  HCO$^{+}$&&  $^{12}$CO & HCO$^{+}$&&  \\
 \hline
1 && 6.60 & 0.15 && - 47.1 & - 49.0 && 3.6 & 1.8  && 2.2 \\
2 && 6.60 & 0.15 && - 46.1 & - 45.3 && 3.6 & 3.7  && 2.2 \\
3 && 6.50 & 0.15 && - 46.4 & - 46.7 && 4.3 & 5.6  && 2.3 \\
4 && 6.90 & 0.23 && - 48.2 & - 47.2 && 4.3 & 3.7  && 3.3 \\
5 && 7.80 & 0.23 && - 48.2 & - 46.7 && 4.3 & 4.7  && 2.9 \\
6 && 7.20 & 0.23 && - 48.2 & - 49.5 && 4.3 & 4.7  && 3.1 \\
7 && 6.30 & 0.23 && - 48.2 & - 49.9 && 5.1 & 4.7  && 3.6 \\
8 && 5.60 & 0.40 && - 48.2 & - 50.0 && 4.3 & 3.7  && 7.1 \\
9 && 6.70 & 0.30 && - 48.6 & - 49.5 && 4.3 & 3.7  && 4.4 \\ 
\hline
\end{tabular}
\label{obstab}
\end{table*}

\begin{figure*}[tt]
\centering
\includegraphics[width=12cm]{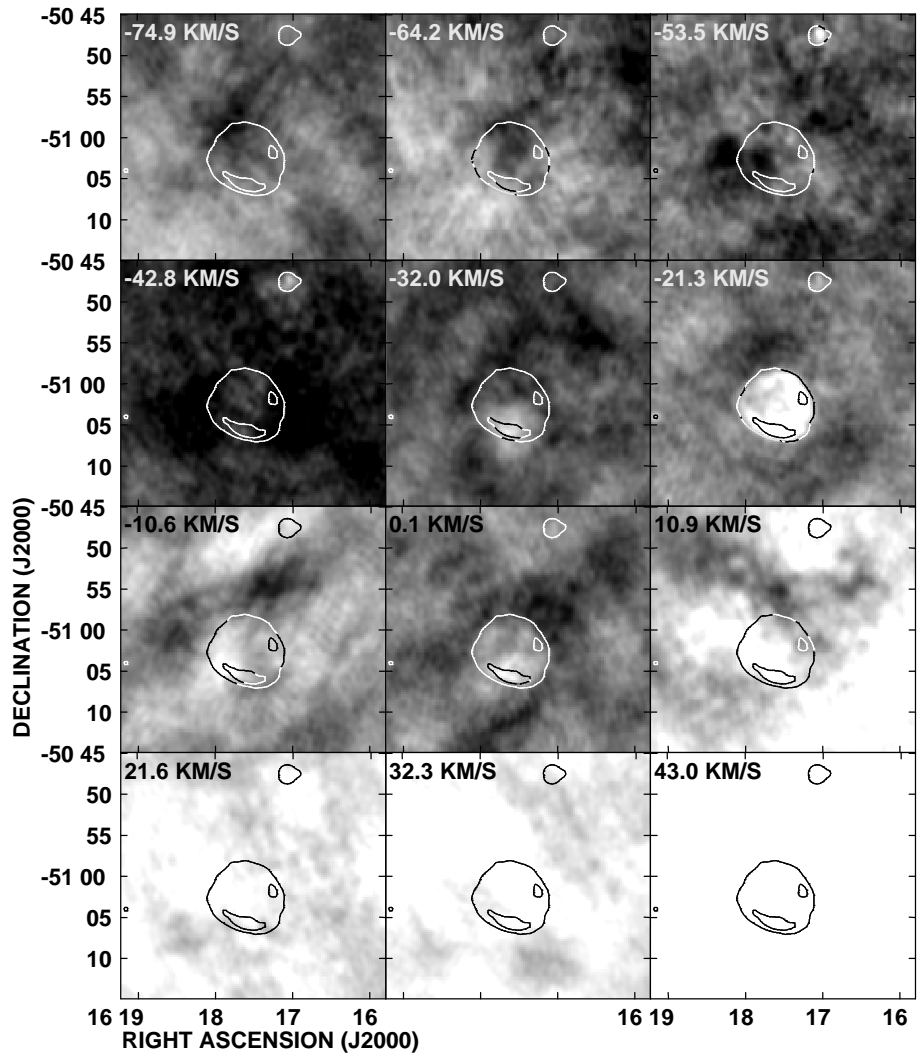}
\caption{HI images between $-$75 and $+$43 \k, each one integrated over $\sim$ 
10 \k. The velocity shown in the upper left corner of each image corresponds to 
the first channel of integration. The grayscale ranges from 400 to 1200 K \k. 
A few contours of the radio continuum of RCW 103 are included for comparison.}
\end{figure*}

\begin{figure*}
\centering
\includegraphics[width=10cm]{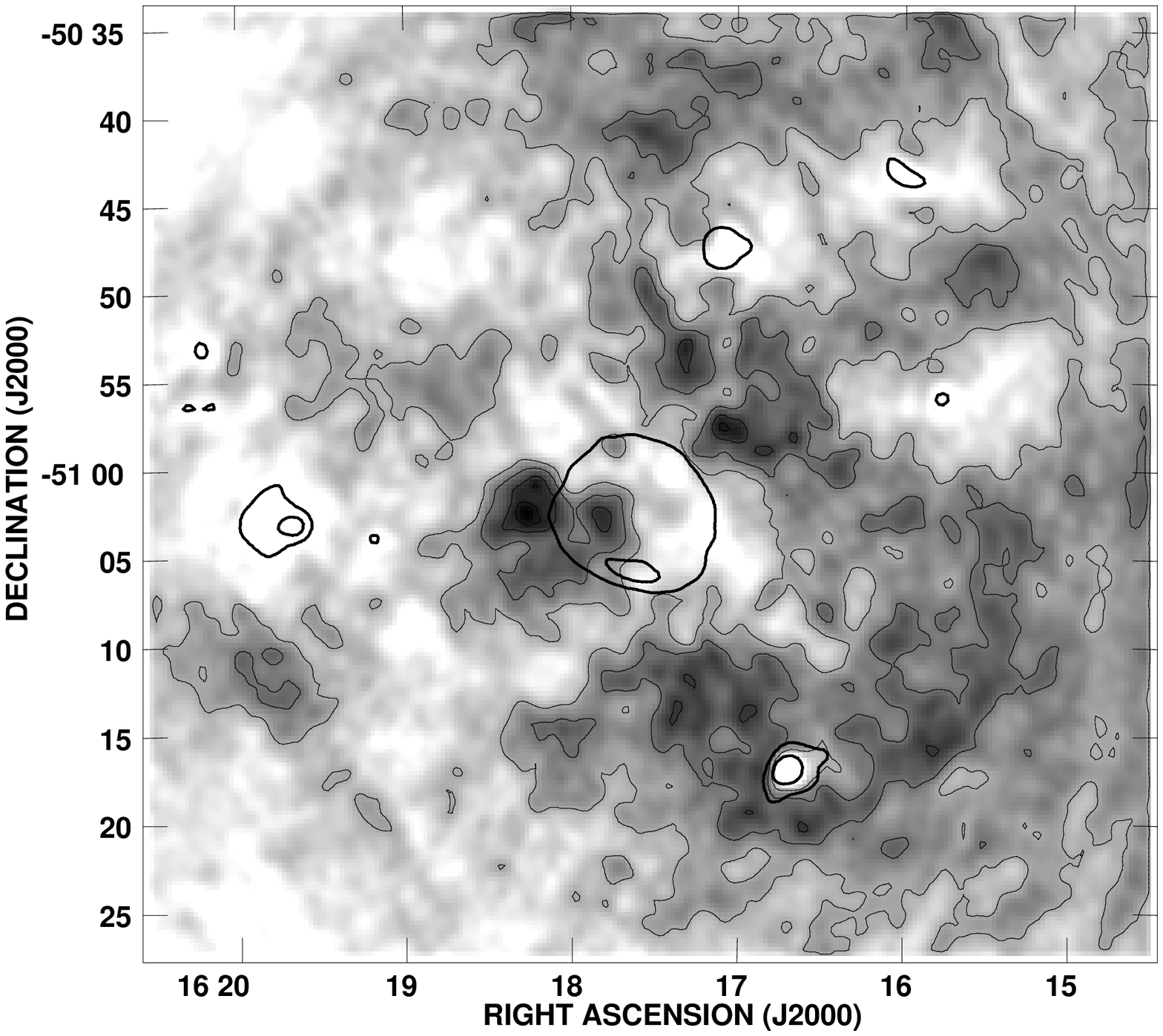}
\caption{Total HI $\lambda$ 21 cm emission integrated between $-$52 and $-$43 
km~s$^{-1}$. The grayscale varies between 650 and 1100 K \k. Contours at 800, 
900, 1000 and 1050 K \ks are included to emphasize the morphology of the 
atomic gas. A few contours from the radio continuum image of RCW 103 are 
shown for comparison. The rms noise level of the image is $\sim$ 47 K \ks and 
the resolution is $50''\times~50''$.}
\end{figure*}

\subsection{Atomic Gas}

Figure 4 shows the distribution of HI in grayscale from {\it v} = $-$75 \k~to {\it v} = 
$+$43 \k~integrated in 10 \ks steps. The contours correspond to the radio 
continuum emission of RCW 103. These velocity slices reveal the patchy 
appearance of the cold gas in the surroundings of RCW 103. The hole in the 
emission evident in the map at $-21.3$ km~s$^{-1}$, corresponds to absorption 
by the Sagitarius-Carina Galactic arm, whose distance is $\sim$1.5 kpc. 
Since HCO$^{+}$ is observed only between $-$52 and $-$43 km~s$^{-1}$, 
Figure 5 shows the HI $\lambda$ 21 cm emission integrated over this velocity 
range to determine if there is an atomic gas counterpart to 
the molecular feature. There is a hint of a shell of atomic gas open to the NE 
of the SNR, and a structure partially overlaping the remnant towards the East.
Assuming that the gas is optically thin and using the relation 
${\rm N}({\rm HI}) = 1.82\times 10^{18} \int{T_{B}~dv}$, where T$_{B}$ is the 
brightness temperature, the HI column density integrated in the above velocity 
range is estimated to be $ N(HI) \sim 1.6\times 10^{21}~\rm cm^{-2}$ for the 
shell and the East clump. Based on these values and assuming a distance of 3.3 
kpc for RCW 103 (from section 4.2), we calculate an HI mass $\sim$ 800 \msol~
for the East clump and aproximately 4400 \msol~for the open shell. Curiously, 
no atomic gas is detected close to the southern border of RCW 103 down to a 
detection level of 3$\sigma$. This fact is discussed later.

\section{ Discussion }

\subsection{Molecular Gas}

From Figure 3, it is clear that the HCO$^{+}$ intensity decreases from outside 
to the interior of the SNR shell, with the emission being weakest for the 
pointings 1 to 3. The coincidence between the shock front position and the 
change in the HCO$^{+}$ intensity strongly suggests the existence of a physical 
interaction between the SNR and the molecular gas. To investigate if such a 
change is the consequence of the passage of the shock front or is simply due to 
a decrease in the ambient molecular density, it is necessary to compare the 
HCO$^{+}$ emission with that of another molecule that remains unnaffected by 
the shock passage. For that purpose, we compared the HCO$^{+}$ emssion with 
that of $^{12}$CO. The use of $^{12}$CO as a reference molecule may be 
problematic because the excitation conditions between $^{12}$CO and a high 
dipole moment species like HCO$^{+}$ could be different. However, assuming that 
$^{12}$CO and HCO$^{+}$ are uniformly mixed, the comparison would be meaningful.
We find that the T$^{\rm HCO^{+}}$/T$^{^{12}\rm CO}$ ratio is lower for 
pointings 1 to 3 than for those that lie well beyond the SNR shock front.
For the inner pointings, the lower values in the intensity of the HCO$^{+}$ 
emission are probably indicative of shock depletion of this molecule. Thus, 
our results confirm that the SNR shock affects the chemistry of the molecular 
gas, but do not support models that predict HCO$^{+}$  enhancement by the 
passage of a shock. 

However, the decrement is not as low as that predicted by the Iglesias \& Silk 
(1978) model, which produces a decrease of two orders of magnitude in the 
HCO$^{+}$ abundance behind a shock front, on time scales of 1000 yr and for
shock velocities of $\sim$ 10 \k. A short interaction time or a shock front with
different velocity could explain the smaller decrease observed in the HCO$^{+}$ 
emission behind the shock, as compared to the theoretical predictions. 
Nevertheless, to determine the HCO$^{+}$ abundance accurately, it would be 
useful to observe a higher dipole moment molecule, such as HCN, rather than 
$^{12}$CO \citep[see][]{ziurys} for the reference molecule.

\subsection{Kinematical Distance}

Previous distance studies suggested a range of possible values between 3.1 and 
4.6 kpc for RCW 103 \citep{reyno04}. The velocity at which the HCO$^{+}$ 
emission is produced, coincident with that of the most intense $^{12}$CO peak, 
allows us to constrain the systemic velocity of RCW 103 to about $-48$ \k. This
velocity corresponds to  a distance of $\sim$ 3.3 kpc using the Galactic 
rotation model of \citet{fich}.

\subsection{Atomic Gas}

From Figure 4 and Figure 5, it is clear that on the southern side of RCW 103, 
where the remnant may be interacting with a molecular cloud, we do not 
observe dense HI emission adjacent to the SN shock. It is possible that the 
lack of HI emission in this region is not due to the absence of atomic 
hydrogen, but to self-absorption processes, as can be seen in Tycho's SNR. 
Proper motion measurements \citep{reyno97} showed that the lowest shock 
expansion velocities around the outer shell of Tycho's SNR are found to the 
East, and it was suggested that a higher ambient density was decelerating the 
expansion in this direction. However, a subsequent study \citep{reyno99} 
detected faint HI emission in this region. The observations can be explained 
if a cooler HI component is self-absorbing a warmer one. \citet{lee} confirmed 
this hypothesis with detection of the molecular emission counterpart of the HI 
self-absorption. The presence of ambient, unshocked molecular gas is indicative 
of low temperatures. A similar scenario for RCW 103 also is consistent with 
Burton \& Spyromilio's (1993) model, in which a reverse shock is propagating 
back into the swept-up atomic gas following the interaction of the blast wave 
with a dense molecular cloud. The blast wave is propagating at 15 \ks and  
the molecular cloud is located ahead of the neutral hydrogen mass. Moreover, 
since the reverse shock has already overrun the HI cloud, the shocked 
gas will be much warmer than the unshocked gas which is mixed with the 
molecular ambient cloud, thus producing the observed self-absorption. In this 
context, the eastern HI clump could be part of an extended structure which 
has a uniform temperature and for which the emission is not self-absorbed. A 
sketch of the possible gas distribution is shown in Figure 6. The cool 
molecular component appears closer to us than the warmer atomic gas and the 
supernova shock front is currently running into the molecular cloud.

In \citet{reyno04}, the absorption spectrum towards RCW 103 was analyzed by 
applying different filtering techniques. In their Figure 2, an absorption 
feature that was seen well above the noise ($> 6 \sigma$) at $+$34 km s$^{-1}$, 
disappeared after deeper filtering and hence, was considered not to be 
caused by absorption against the continuum emission from RCW 103. To review
a possible association with RCW 103, a map of the HI distribution at this 
velocity was produced.

Figure 7 shows the HI distribution integrated within 4 km s$^{-1}$ around $+$34 
km s$^{-1}$. The neutral gas depicts an open shell-like structure well-matched
to the SNR ring. Since the molecular data presented here and the HI absorption 
study in Reynoso et al. (2004) have shown that the systemic velocity of RCW 103 
is $-48$ km s$^{-1}$, it is highly unlikely that this HI shell at $+$34 km 
s$^{-1}$ (which corresponds to a kinematic distance of 19 kpc using the Galactic
rotation model of Fich et al. 1989) has been driven by the SN shock front. 

\begin{figure}
\centering
\includegraphics[width=7cm]{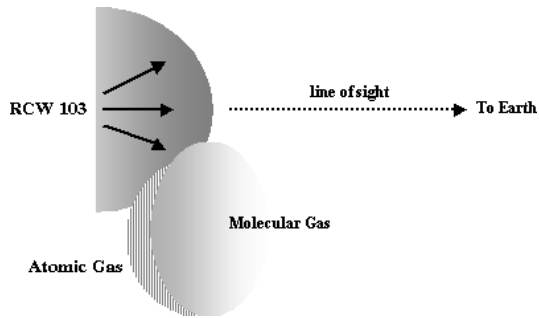}
\caption{A schematic model of the RCW 103 interaction with a molecular cloud to
the South.}
\end{figure}

Although there is a good morphological match between the radio continuum
extent and the HI feature, it is very difficult to explain any relationship
between the SNR and this HI structure. The most likely explanation is chance
juxtaposition and no relationship.

\begin{figure}
\centering
\includegraphics[width=7cm]{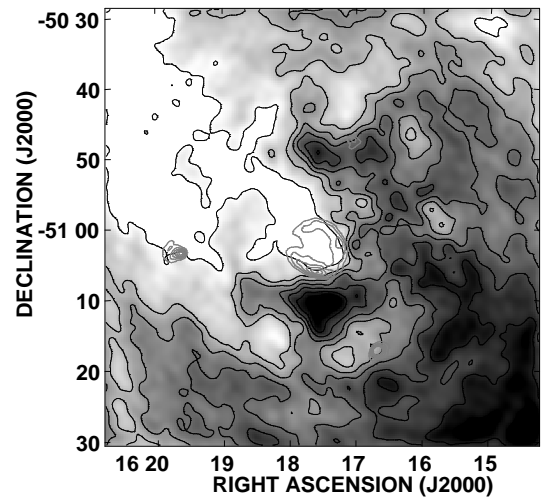}
\caption{HI emission integrated within 4 km s$^{-1}$ around +34 km s$^{-1}$.  
The grayscale varies between 30 and 70 K km s$^{-1}$. The HI contour levels 
(in black) are 30, 40, 50, 60 and 65 K km s$^{-1}$. A few contours (in grey) 
representing the radio continuum of RCW 103 are shown for comparison. }
\end{figure}

\section{Summary}

We have investigated the interstellar medium around the SNR RCW 103. The 
present $^{12}$CO J=1-0 and HCO$^{+}$ J=1-0 observations provide new evidence 
for an interaction between the SNR and a molecular cloud. We also investigated 
the surrounding atomic gas based on the $\lambda$ 21 cm HI line. The main 
results can be summarized as follows:

(a) From J=1-0 $^{12}$CO and HCO$^{+}$ observations towards the southern side 
of RCW 103, we find that the HCO$^{+}$ spectra have only a single component 
which has at the same kinematical velocity of $\sim -48$ \ks as the brightest
feature observed in the $^{12}$CO spectrum. 

(b) The T$^{\rm HCO^{+}}$/T$^{^{12}\rm CO}$ ratio at this velocity range 
has lower values for the interior pointings (1 to 3) than for those that are
well away from the SNR shock front. The low ratio values for the innner 
positions may be due to HCO$^{+}$ depletion caused by the remnant shock front 
colliding with the molecular gas.

(c) The molecular observations allows us to constrain the systemic velocity of 
RCW 103 to about $-48$ \k. This velocity corresponds to a distance of $\sim$ 
3.3 kpc for the remnant, using the Galactic rotation model of \citet{fich}. 

(d) We do not detect HI emission on the southern side of the remnant at
$-48$ km~s$^{-1}$, where molecular gas is observed. We suggest that this
result is due to self-absorption processes, similar to those observed in 
Tycho's SNR. 

\section*{Acknowledgments}
We thank Michael Burton for help with the molecular line proposal.  S.P. thanks 
Facundo Albacete for a fruitful discussion about the X-ray image. S.P. is a 
fellow of CONICET, Argentina. E. R. and G. D. are members of the {\it Carrera 
del Investigador Cient\'\i fico}, CONICET, Argentina. During part of this work, 
E. M. R. was a visiting scholar at the University of Sydney.
This research was partially funded through the Australian Research Council, 
by the UBACYT Grant A055 and by ANPCyT-PICT04-14018 (Argentina). 
The ATCA is funded by the Commonwealth of Australia for operation as a
National Facility by CSIRO.

\end{document}